\documentclass[prl,showpacs,twocolumn]{revtex4}
\usepackage{mathrsfs}
\usepackage{amsmath}
\usepackage{amssymb}
\usepackage{revsymb}
\usepackage{graphicx}
\usepackage{mathrsfs}
\begin{document}
\title{Strong signatures of radiation reaction below the radiation-dominated regime}
\author{A. \surname{Di Piazza}}
\email{dipiazza@mpi-hd.mpg.de}
\author{K. Z. \surname{Hatsagortsyan}}
\email{k.hatsagortsyan@mpi-hd.mpg.de}
\author{C. H. \surname{Keitel}}
\email{keitel@mpi-hd.mpg.de}
\affiliation{Max-Planck-Institut f\"ur Kernphysik, Saupfercheckweg 1, D-69117 Heidelberg, Germany}

\date{\today}

\begin{abstract}
The influence of radiation reaction (RR) on multiphoton Thomson scattering by an electron colliding head-on with a strong laser beam is investigated in a new regime, in which the momentum transferred on average to the electron by the laser pulse approximately compensates the one initially prepared. This equilibrium is shown to be far more sensitive to the influence of RR than previously studied scenarios. As a consequence RR can be experimentally investigated with currently available laser systems and the underlying widely discussed theoretical equations become testable for the first time.

 
\end{abstract}
 
\maketitle

An accelerated electric charge emits electromagnetic radiation and, in turn, this emission modifies the motion of the charge itself. In classical electrodynamics the Lorentz-Abraham-Dirac (LAD) equation is the covariant equation of motion of a charge that accounts self-consistently for the effects of radiation reaction (RR) as an additional four-force \cite{LAD} (see \cite{Books_RR} for a recent review). It is well known that the LAD equation presents physically unsatisfactory features like the existence of runaway solutions in which the charge acceleration increases exponentially even without an external field or the violation of the causality principle. However, these problems are not genuine in the realm of classical electrodynamics. In fact, in the non-relativistic limit the RR force on the charge is always much smaller than the Lorentz force and this is also the case in the relativistic regime but in the instantaneous rest frame of the charge \cite{Landau_b_2_1975,Footnote_1}. By employing this property, one can consistently perform a perturbative iteration in the RR terms in the LAD equation and obtain the so-called Landau-Lifshitz (LL) equation which is not plagued by the mentioned problems \cite{Landau_b_2_1975}. Recently, it has been shown that the solutions of the LL equation cover the entire \emph{physical} sub-manifold of the solutions of the LAD equation \cite{Spohn_2000,Rohrlich_2002}. Noticeably, unlike in the non-relativistic case, a regime can arise in the ultra-relativistic case where the RR force in the LL equation becomes comparable with the Lorentz force in the laboratory frame, while being much smaller in the instantaneous rest frame of the charge \cite{Landau_b_2_1975}. This is the so-called radiation dominated regime (RDR) in which the charge dynamics is significantly modified due to RR. 

An experimental confirmation of the LL equation (and of the LAD equation, as well) is still missing. Presently available lasers produce intense electromagnetic fields and can represent a unique tool to investigate experimentally RR. Nowadays, the record intensity of $2\times 10^{22}\;\text{W/cm$^2$}$ has been obtained \cite{I_max} and the upcoming petawatt lasers aim at intensities of the order of $10^{23}\;\text{W/cm$^2$}$ \cite{Norby_2005}. In the near future, intensities of the order of $10^{24}\text{-}10^{26}\;\text{W/cm$^2$}$ are envisaged at the Extreme-Light-Infrastructure (ELI) \cite{ELI_Laser}. 

The RDR in a laser field is determined by comparing the energy loss of the charge in one laser period due to RR with its initial energy \cite{Koga_2005}. For an electron (charge $-e<0$ and mass $m$) with an initial Lorentz-factor $\gamma_0$ colliding head-on with a laser field with angular frequency $\omega_0$ and electric field amplitude $E_0$ the parameter characterizing this regime reads
\begin{equation}
\label{R}
R=\frac{2}{3}\alpha\frac{\omega_0}{m}\gamma_0(1+\beta_0)\xi^2,
\end{equation}
and the RDR arises when $R\approx 1$. Here, $\alpha=e^2$ is the fine-structure constant, $\beta_0=\sqrt{\gamma_0^2-1}/\gamma_0$ and $\xi=eE_0/m\omega_0$ (units with $\hbar=c=1$ are used throughout). 

In the present Letter we show that in the specific domain of parameters $2\gamma_0 \gtrapprox \xi$ a new radiation regime below the RDR ($R\ll 1$) can be attained in which the impact of RR on the electron motion and on the emission spectra is qualitatively and quantitatively significant. This occurs because in this regime the change of the electron momentum along the laser propagation direction (longitudinal momentum) due to RR in one laser period is of the order of the longitudinal momentum itself in the laser field and the electron, initially counterpropagating with the laser beam, undergoes a reflection only due to RR. The condition characterizing this regime for an ultra-relativistic electron ($\gamma_0\gg 1$) and a short, strong ($\xi\gg 1$) laser pulse is (see derivation below)
\begin{equation}
\label{cond_domin2}
R\gtrsim \frac{4\gamma_0^2-\xi^2}{2\xi^2}>0,
\end{equation}
which is much less restrictive than $R\approx 1$ as it can also hold at $R\ll 1$ if $4\gamma_0^2 \approx \xi^2$ (the requirement $4\gamma_0^2-\xi^2>0$ will be clarified below). The experimental feasibility of this regime with an all-optical setup by employing a laser system with an intensity of the same order of those presently available is demonstrated in principle by means of a realistic numerical example. This can represent the first experimental test of the validity of the LL equation.

RR effects in Thomson scattering have been investigated in \cite{Koga_2005} at $R\sim 1$ and in \cite{Hartemann_1996,Keitel_1998,Cang_2006} at $R\ll 1$. However, physical situations have solely been considered in which either $\xi\gg \gamma_0$ \cite{Keitel_1998} or $\xi\ll \gamma_0$ \cite{Koga_2005,Hartemann_1996,Cang_2006} rendering condition (\ref{cond_domin2}) impossible to be fulfilled. As a result, the effects of RR on the radiation spectrum \cite{Koga_2005,Hartemann_1996,Keitel_1998}, on the electron trajectory \cite{Keitel_1998} and on the temporal structure of the emitted radiation \cite{Cang_2006} are only quantitative and rather small at $R\ll 1$. 

The LL equation for an electron in the presence of an external electromagnetic field $F^{\mu\nu}(x)$ ($x$ indicates the four space-time coordinates, i. e. $x^{\mu}=(t,\mathbf{r})$) is \cite{Landau_b_2_1975}
\begin{equation}
\label{LL_eq}
\begin{split}
&m\frac{d u^{\mu}}{ds}=-eF^{\mu\nu}u_{\nu}-\frac{2}{3}\alpha\bigg[\frac{e}{m}(\partial_{\alpha}F^{\mu\nu})u^{\alpha}u_{\nu}\\
&+\frac{e^2}{m^2}F^{\mu\nu}F_{\alpha\nu}u^{\alpha}-\frac{e^2}{m^2}(F^{\alpha\nu}u_{\nu})(F_{\alpha\lambda}u^{\lambda})u^{\mu}\bigg].
\end{split}
\end{equation}
Here $s$ is the electron proper time, $u^{\mu}=dx^{\mu}/ds$ the electron four-velocity with metric $g^{\mu\nu}=\text{diag}\{+1,-1,-1,-1\}$, and the terms proportional to $\alpha$ arise due to RR. For a qualitative investigation of the new radiation regime characterized by the condition (\ref{cond_domin2}), we first employ the exact analytical solution of the LL equation in the presence of an arbitrary plane wave \cite{Di_Piazza_2008}. If the electron has initial four-velocity $u_0^{\mu}=\gamma_0(1,0,-\beta_0,0)$ and the plane wave propagates along the positive $y$ direction with linear polarization along the $z$ direction, central angular frequency $\omega_0$ (wavelength $\lambda_0=2\pi/\omega_0$) and electric field amplitude $E_0$, the solution depends only on the laser phase $\phi=\omega_0(t-y)$ and reads
\begin{equation}
\label{u}
u^{\mu}(\phi)=\frac{1}{h(\phi)}
\begin{pmatrix}
\gamma_0+\frac{\omega_0}{2m\eta_0}[h^2(\phi)-1+\xi^2\mathcal{I}^2(\phi)]\\
0\\
-\beta_0\gamma_0+\frac{\omega_0}{2m\eta_0}[h^2(\phi)-1+\xi^2\mathcal{I}^2(\phi)]\\
-\xi \mathcal{I}(\phi)
\end{pmatrix}.
\end{equation}
In this expression $\eta_0=\omega_0\gamma_0(1+\beta_0)/m$ and 
\begin{align} 
\label{h}
h(\phi)&=1+R
\int_{\phi_0}^{\phi}d\zeta\psi^2(\zeta),\\
\label{I}
\mathcal{I}(\phi)&=\int_{\phi_0}^{\phi}d\zeta\left[h(\zeta)\psi(\zeta)+
\frac{R}{\xi^2}\frac{d\psi(\zeta)}{d\zeta}\right],
\end{align}
where $\phi_0$ is the initial phase at which $u^{\mu}(\phi_0)=u_0^{\mu}$ and the function $\psi(\phi)$ is defined through the plane wave electric field $\mathbf{E}(\phi)$ as $\mathbf{E}(\phi)=E_0\psi(\phi)\hat{\mathbf{z}}$. The terms due to RR in the solution (\ref{u})-(\ref{I}) scale with the parameter $R$ given in Eq. (\ref{R}). Therefore, one is led to think that in order to have significant effects of RR, one has to consider the $R\approx 1$ regime. However, Eqs. (\ref{u})-(\ref{I}) indicate that even if $R\ll 1$ there are situations in which the longitudinal velocity of the electron (the $y$ component) can change sign only if RR effects are taken into account. From a mathematical point of view this is the case essentially because the function $h(\phi)$ (that is also contained in $\mathcal{I}(\phi)$) is larger when the RR terms are included. In turn, this has also a clear physical interpretation: in the presence of RR the electron constantly looses energy due to electromagnetic radiation and then it is more easily reflected when traveling through the laser field. In order to give a qualitative condition for the electron being reflected only in the presence of RR, we consider the relevant situation where $\gamma_0,\,\xi\gg 1$ and the pulse length is not much larger than the laser period. We can separate the terms arising from RR from those present also in the absence of RR in the functions $h(\phi)$ and $\mathcal{I}(\phi)$ as: $h(\phi)=1+Rf(\phi)$ with $f(\phi)=\int_{\phi_0}^{\phi}d\zeta\psi^2(\zeta)\sim 1$ and $\mathcal{I}(\phi)\approx\mathcal{I}_0(\phi)+Rg(\phi)$ with $\mathcal{I}_0(\phi)=\int_{\phi_0}^{\phi}d\zeta\psi(\zeta)\sim 1$ and $g(\phi)=\int_{\phi_0}^{\phi}d\zeta f(\zeta)\psi(\zeta)\sim 1$. By assuming that $R\ll 1$, the condition $u_y(\phi)>0$ reads $2R\xi^2g(\phi)>4\gamma_0^2-\xi^2\mathcal{I}_0^2(\phi)>0$  which, more qualitatively, gives Eq. (\ref{cond_domin2}). A more intuitive way of obtaining this equation is to require the variation of the longitudinal momentum of the electron in one laser period, estimated as $mR$ from Eq. (\ref{LL_eq}), to be comparable with the longitudinal momentum itself in the laser field. Note that the condition $4\gamma_0^2-\xi^2\mathcal{I}_0^2(\phi)>0$ ensures that the electron is not reflected without RR. Since the electron is assumed to be ultrarelativistic, it will emit mainly along its velocity \cite{Landau_b_2_1975}. Therefore, by considering spherical coordinates $(r,\vartheta,\varphi)$ with the $y$ axis as the polar axis, we can predict that in the present regime the electron emits almost entirely for $\vartheta\ge 90^{\circ}$ in the absence of RR while it also emits significantly at $\vartheta < 90^{\circ}$ in the presence of RR.

In order to support our theoretical predictions in the simplified situation discussed above, we consider below the more realistic situation in which the electron is driven by a focused laser field with the above employed amplitude $E_0$ and central angular frequency $\omega_0$ and with spot radius $\sigma_0$. The laser beam has been modeled as a Gaussian paraxial beam up to terms linear in the small parameter $\epsilon=\lambda_0/\pi\sigma_0$, i. e. including the longitudinal electric and magnetic fields (see, e. g., Eqs. (1)-(9) in \cite{Salamin_2002}), with a $\sin^2$-time envelope. We consider an electron with an initial energy of $40\;\text{MeV}$ and a Ti-Sapphire laser field with $\omega_0=1.55\;\text{eV}$, an intensity of $I_0=5\times 10^{22}\;\text{W/cm$^2$}$ ($\xi\approx 150$), a pulse duration of $27\;\text{fs}$ and $\sigma_0=2.5\;\text{$\mu$m}$ ($\epsilon=0.1$). The laser spot is centered in the coordinates' origin which coincides with the initial electron position. We have ensured that all the conditions concerning the validity of classical electrodynamics and of Eq. (\ref{LL_eq}) are fulfilled with the above numerical parameters. In Fig. 1 we show the electron trajectory calculated with the above parameters and (a)) by removing and (b)) by keeping the RR terms in Eq. (\ref{LL_eq}).
\begin{figure}
\begin{center}
\includegraphics[width=6cm]{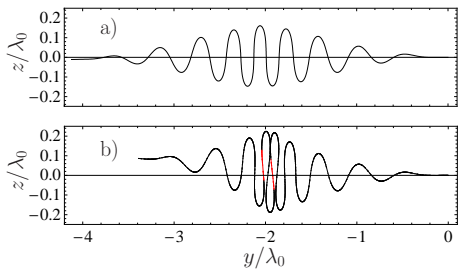}
\end{center}
\caption{The electron trajectories in units of the laser wavelength $\lambda_0$ calculated (a)) by removing  and (b)) by keeping the RR terms in Eq. (\ref{LL_eq}). The initial electron energy is $40\;\text{MeV}$, the laser field intensity $I_0=5\times 10^{22}\;\text{W/cm$^2$}$, the wavelength $\lambda_0=0.8\;\text{$\mu$m}$, the pulse duration $27\;\text{fs}$ and the waist size $\sigma_0=2.5\;\text{$\mu$m}$. The red (gray) portions of the trajectory are those in which the longitudinal velocity of the electron is positive.}
\end{figure}
As expected, in the presence of RR the electron has a positive longitudinal velocity for a significant portion of the trajectory (indicated in red (gray) in Fig. 1 b)). Since the electron is ultra-relativistic, the fact that its velocity also points to directions with $\vartheta < 90^{\circ}$ has a significant impact on the angular distribution of the emitted radiation. We have calculated the angle resolved spectral energy $dW/d\omega d\Omega$ with $d\Omega=\sin\vartheta d\vartheta d\varphi$ at different angles $\vartheta$ and at $\varphi=0$ and $\varphi=180^{\circ}$ (corresponding to the $y\text{-}z$ plane where the electron mainly moves) by employing Eq. (14.60) in \cite{Jackson_b_1975}. The results in the region $\varphi=180^{\circ}$ and $\vartheta\in [70^{\circ},110^{\circ}]$ without and with RR are shown in Fig. \ref{Fig_2} a) and b), respectively. We focus only on this angular region because in both cases at smaller $\vartheta$ there is no significant emission and at larger $\vartheta$ the spectra are similar to those at $\vartheta>90^{\circ}$. The black lines indicate the cutoff position $\omega_c/\omega_0$ at each observation angle which is estimated by using the well-known expression $\omega_c/\omega_0\approx \displaystyle\max_i\{3\gamma^3(t_i)/\rho(t_i)\}= 3\gamma^3(t_c)/\rho(t_c)$ \cite{Jackson_b_1975}, where $\rho(t)$ is the curvature of the trajectory as a function of time and $t_i$ are the values of time at which the electron velocity points along the observation direction. 
We note from the figure that the estimated cutoff is in qualitatively good agreement with the numerical results. At angles $\vartheta > 90^{\circ}$ the spectra with and without RR are qualitatively similar, the main differences being, as expected, a lower overall amplitude and a lower cutoff position in the case in which RR effects are included. However, in the other angular region $\vartheta < 90^{\circ}$, there is essentially no emission without RR while with RR the emission extends significantly down to $\vartheta\approx 75^{\circ}$. As an example we consider the case $\vartheta=80^{\circ}$ and we indicate as $\Omega_0$ the direction with $\vartheta=80^{\circ}$ and $\varphi=180^{\circ}$.
\begin{figure}
\begin{center}
\includegraphics[width=6.5cm]{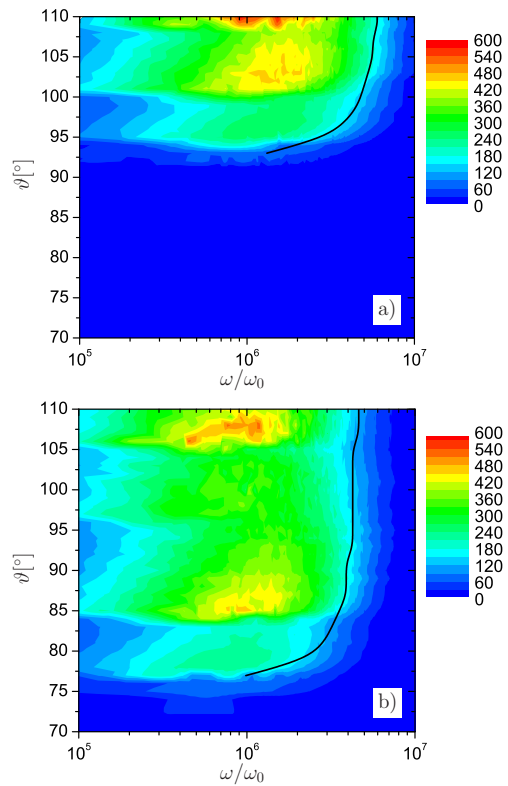}
\end{center}
\caption{Angle resolved spectral energy $dW/d\omega d\Omega$ in $\text{sr}^{-1}$ emitted by the electron at $\varphi=180^{\circ}$ without (a)) and with (b)) RR. The electron and the laser field parameters are the same as in Fig. 1.}
\label{Fig_2}
\end{figure}
Only a few tenths of harmonics are emitted without RR with a peak amplitude of the order of $10^{-2}\;\text{sr$^{-1}$}$ (not shown in Fig. 2), while harmonics up to a few million are significantly emitted ($dW/d\omega d\Omega\sim 10^2\;\text{sr$^{-1}$}$) with RR, corresponding to energies $\omega$ of the order of MeV. Finally, it is worth stressing that in the present numerical example the parameter $R$ is very small: $R=5\times 10^{-2}$. Therefore, as already mentioned, we are far below the RDR. 

\begin{figure}
\begin{center}
\includegraphics[width=6cm]{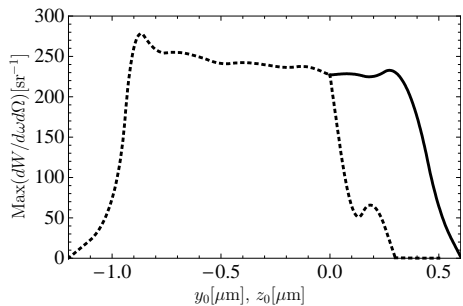}
\end{center}
\caption{Maximum value of the angle resolved spectral energy along $\Omega_0$ with RR as a function of the initial electron coordinate $y_0$ (continuous line) and $z_0$ (dotted line). The electron and the laser field parameters are the same as in Fig. 1.}
\label{Fig_3}
\end{figure}

For the sake of a measurable photon yield, an electron beam should be employed in an experimental implementation rather than a single electron. Therefore, the possible influence of multi-particle effects should be estimated. Laser-plasma accelerators (LPAs) \cite{Dream_beams} can provide electron beams with energies of the order of $40\;\text{MeV}$ like those required here and are very suitable in the present situation. They allow for an all-optical setup and a very good beam-beam overlapping as the dimensions of the electron beams generated by LPAs are of the same order as those of the strong laser beam. On one hand, the maximal value of the space-charge force can be estimated as $F_C\sim 2\pi \alpha nw_0$, with electron density $n$ and beam waist size $w_0\sim \sigma_0$, while the RR force yields $F_R\sim\alpha^2\gamma^2E^2_0/m^2$. Since the beams obtained with LPAs contain up to $10^9$ electrons \cite{Malka_2008}, by assuming a volume of the electron beam of about $1000\;\text{$\mu$m$^3$}$, we obtain that the space-charge field is negligible with respect to the RR force as $F_C/F_R\sim 10^{-3}$. On the other hand, in the relevant part of the energy spectrum ($\omega/\omega_0>10^5$), the wavelengths ($\lambda <10^{-2}\,\text{nm}$) are much less than the mean distance between the electrons in the beam which is about $10\;\text{nm}$. Consequently, the collective effects in the radiation are not significant and different electrons in the beam contribute incoherently to this part of the spectrum. Further, we investigate the effect of incoherent averaging of the radiation spectrum over the initial conditions of the electron. The variation of the initial electron coordinate along the $x$ direction within the beam waist size causes a slight modification of the direction of main emission along $\varphi$ without a significant decrease in intensity. In addition, changing the initial position along the laser propagation direction ($y$) or the laser polarization direction ($z$) can alter significantly the electron trajectory. As an example, in Fig. \ref{Fig_3} we show the maximum value of the electron spectrum with RR emitted along $\Omega_0$, as a function of $y_0$ (continuous line) and of $z_0$ (dotted line).
We observe that the reduction of the spectral amplitude is not significant when the variation of the coordinate along the $y$-axis is less than $0.5\;\text{$\mu$m}$ and along the $z$-axis less than $0.5\text{-}1\;\text{$\mu$m}$. We have not considered the case $y_0<0$ which implies the electron to be initially already in the laser beam. The asymmetry in the case of $z_0$ depends on the carrier envelope phase (CEP) of the laser pulse but the electron reflection with RR occurs independently of the CEP of the laser beam. Moreover, we have ensured that the energy spread ($\sim 1\;\%$) and the beam divergence ($\sim 5\;\text{mrad}$) attainable by LPAs \cite{Malka_2008} do not modify the electron trajectory significantly. Also, a change in the laser intensity of $\pm 10\;\%$ accounting for experimental uncertaincies \cite{I_max} results into a shift of the minimum angle of emission of about $\mp 5^{\circ}$ both with and without RR that does not affect our conclusions.

Finally, by integrating the electron spectrum in Fig. \ref{Fig_2} b) with respect to $\omega$ at $\vartheta=80^{\circ}$, and by estimating $\Delta\Omega\sim 1/\gamma(t_c)\approx 7\times 10^{-3}\;\text{sr}$ we find that a total energy of about $1.9\;\text{MeV}$ is emitted along $\Omega_0$. Since the maximum emission is at $1.3\;\text{MeV}$, we expect approximately per shot one photon to be emitted along $\Omega_0$. Germanium detectors are suitable for this energy range and nowadays they reach an efficiency of about $10^{-3}$ \cite{Lee_2003}. Moreover, from the analysis in Fig. \ref{Fig_3} we can roughly estimate that in this case $1\;\%$ of the electrons in the bunch will contribute to the considered effect. Therefore, by employing an electron beam with $10^9$ electrons we would expect about $10^4$ photons registered per shot.

In conclusion, we have demonstrated that RR as predicted by the LL equation can have a dramatic impact on the electron spectra even if the RR parameter $R$ is much smaller than unity. This result has been obtained in a new physical regime in which the electron is temporarily reflected due to RR when it travels through a laser beam. We have shown that the effect predicted is measurable in principle by employing presently available detection systems and electron beams together with laser systems with intensities a little larger but of the same order than those already available. This could result in the first experimental test of the LL equation.
%
%


\begin{thebibliography}{99}
\bibitem{LAD} M. Abraham, \textit{Theorie der Elektrizit\"{a}t, Vol. II}, Teubner, Leipzig (1905); H. A. Lorentz, \textit{The Theory of Electrons}, Teubner, Leipzig (1909); P. A. M. Dirac, Proc. Roy. Soc. (London) A \textbf{167}, 148 (1938).
\bibitem{Books_RR} H. Spohn, \textit{Dynamics of charged particles and their radiation field}, Cambridge University Press, Cambridge (2004); F. Rohrlich, \textit{Classical Charged Particles}, World Scientific, Singapore (2007).
\bibitem{Landau_b_2_1975} L. D. Landau, and E. M. Lifshitz, \textit{The Classical Theory of Fields}, Elsevier, Oxford (1975), Par. 74-76.
\bibitem{Footnote_1} 
In classical electrodynamics the RR force on a charge, an electron for definiteness, is much smaller than the Lorentz force when the typical wavelength (amplitude) of the external electromagnetic field in the rest frame of the electron is much larger (smaller) than $\alpha\lambda_c$ ($E_{cr}/\alpha$). Here, $\alpha=e^2/\hbar c\approx 1/137$ is the fine-structure constant, $\lambda_c=\hbar/mc$ is the Compton wavelength and $E_{cr}=m^2c^3/\hbar e$ is the critical field of quantum electrodynamics, where $-e<0$ and $m$ are the electron charge and mass respectively \cite{Landau_b_2_1975}. However, quantum effects become important already at external field wavelengths and amplitudes of the order of $\lambda_c$ and $E_{cr}$, respectively \cite{Ritus_Review}. 
\bibitem{Ritus_Review} V. I. Ritus, J. Sov. Laser Res. \textbf{6}, 497 (1985).
\bibitem{Spohn_2000} H. Spohn, Europhys. Lett. \textbf{50}, 287 (2000).
\bibitem{Rohrlich_2002} F. Rohrlich, Phys. Lett. A \textbf{303}, 307 (2002).
\bibitem{I_max} V. Yanovsky \textit{et al.}, Opt. Express {\bf 16}, 2109 (2008).
\bibitem{Norby_2005} J. Norby, Laser Focus World, January 1 (2005).
\bibitem{ELI_Laser} http://www.extreme-light-infrastructure.eu/.
\bibitem{Koga_2005} J. Koga, T. Zh. Esirkepov, and S. V. Bulanov, Phys. Plasmas \textbf{12}, 093106 (2005).
\bibitem{Hartemann_1996} F. V. Hartemann and A. K. Kerman, Phys. Rev. Lett. \textbf{76}, 624 (1996).
\bibitem{Keitel_1998} C. H. Keitel, \textit{et al.}, J. Phys. B \textbf{31}, L75 (1998).
\bibitem{Cang_2006} Yu Cang \textit{et al.}, Phys. Plasmas \textbf{13}, 113106 (2006).
\bibitem{Di_Piazza_2008} A. Di Piazza, Lett. Math. Phys. \textbf{83}, 305 (2008).
\bibitem{Salamin_2002} Y. I. Salamin and C. H. Keitel, Phys. Rev. Lett. \textbf{88}, 095005 (2002).
\bibitem{Jackson_b_1975} J. D. Jackson, \textit{Classical Electrodynamics}, Wiley, New York (1975), Ch. 14.
\bibitem{Dream_beams} S. P. D. Mangles, \textit{et al.}, Nature \textbf{431}, 535 (2004); C. G. R. Geddes, \textit{et al.}, Nature \textbf{431}, 538 (2004); J. Faure, \textit{et al.}, Nature \textbf{431}, 541 (2004).
\bibitem{Malka_2008} V. Malka, \textit{et al.} Nature Phys. \textbf{4}, 447 (2008).
\bibitem{Lee_2003} I. Y. Lee, M. A. Delephanquem and K. Vetter, Rep. Prog. Phys. \textbf{66}, 1095 (2003).
\end{thebibliography}
\end{document}